\documentclass[aps,prl,twocolumn,preprintnumbers,%
superscriptaddress,floatfix]{revtex4}

\usepackage[dvips]{graphics}
\usepackage{amsmath,amsfonts,bm}
\usepackage{psfig}

\begin{document}

\newcommand\<{\langle}
\renewcommand\>{\rangle}
\renewcommand\d{\partial}
\newcommand\LambdaQCD{\Lambda_{\textrm{QCD}}}
\newcommand\tr{\mathop{\mathrm{Tr}}}
\newcommand\+{\dagger}
\newcommand\g{g_5}

\preprint{WM-05-101, INT-PUB 05-02, SLAC-PUB-10965}

\affiliation{Department of Physics, College of William and Mary,
Williamsburg, Virginia 23187-8795, USA}
\affiliation{Stanford Linear Accelerator Center, Stanford University,
Stanford, California 94309, USA}
\affiliation{Institute for Nuclear Theory, University of Washington,
Seattle, Washington 98195-1550, USA}
\affiliation{Department of Physics, University of Illinois, 
Chicago, Illinois 60607-7059, USA}
\title{QCD and a Holographic Model of Hadrons}
\author{Joshua Erlich}
\affiliation{Department of Physics, College of William and Mary,
Williamsburg, Virginia 23187-8795, USA}
\author{Emanuel Katz}
\affiliation{Stanford Linear Accelerator Center, Stanford University,
Stanford, California 94309, USA}
%\email{amikatz@slac.stanford.edu}
\author{Dam T.~Son}
\affiliation{Institute for Nuclear Theory, University of Washington,
Seattle, Washington 98195-1550, USA}
%\email{son@phys.washington.edu}
\author{Mikhail A.~Stephanov}
\affiliation{Department of Physics, University of Illinois, 
Chicago, Illinois 60607-7059, USA}
%\email{misha@uic.edu}

\date{January 2005}

\newcommand\sect[1]{\emph{#1}---}

\begin{abstract}
We propose a five-dimensional framework for modeling
low-energy properties of QCD. 
In the simplest three parameter model we compute 
masses, decay rates and couplings 
of the lightest mesons. The model fits
experimental data to within 10\%.
The framework is a holographic
version of the QCD sum rules, motivated by the 
anti-de Sitter/conformal field theory (AdS/CFT) correspondence.
The model naturally incorporates properties of QCD dictated
by chiral symmetry, which we demonstrate by deriving the
Gell-Mann--Oakes--Renner relationship for the pion mass.
\end{abstract}
\keywords{QCD, AdS-CFT Correspondence}
\pacs{11.25.Tq, %Gauge/string duality
11.10.Kk, %Field theories in dimensions other than four
11.25.Wx, %String and brane phenomenology
12.38.Cy%Summation of perturbation theory
}
\maketitle

\sect{Introduction.}%
QCD has eluded an analytic solution, despite extensive 
efforts applied to this problem in the past 30 years. Recently, 
the gravity/gauge,
or anti-de Sitter/conformal field theory
(AdS/CFT) correspondence~\cite{Maldacena:1997re} 
has revived the hope that QCD can be
reformulated as a solvable string theory.  So far, theories which can
be solved using AdS/CFT techniques differ substantially from QCD, most
notably by the strong coupling in
the ultraviolet (UV) regime and the lack of asymptotic freedom.  
Nevertheless, certain important properties of QCD, such as
confinement and chiral symmetry breaking, are present in many of these
theories, and the gravity/gauge duality provides a new approach to studying
the resulting dynamics.  
An important development in the prototypical example of
${\cal N}=4$ super Yang-Mills (SYM) theory has been the introduction
of fundamental quarks using
probe D7 branes~\cite{Karch:2002sh}.  The mesons that appear in
these theories behave in many ways similarly to the mesons in
QCD~\cite{Kruczenski:2003be,Hong}.  

Inspired by the gravity/gauge duality  we propose the following complementary
approach. Rather than deform the SYM theory to obtain 
QCD \cite{others}, 
we start from QCD and attempt to construct its five-dimensional (5D)
holographic dual.
In this Letter, we present an exploratory study of a simple holographic model of QCD.  
The field content of the 5D theory is chosen
to reproduce holographically the dynamics of chiral symmetry
breaking in QCD, the boundary theory.  
The model has four free parameters, one of which is fixed
by the number of colors; the remaining three parameters can be fitted using
three well-measured observables, e.g., 
the $\rho$ meson mass, the pion mass, 
and the pion decay constant.  The model then
predicts other low-energy hadronic observables with
surprisingly good accuracy.  

Such an approach is similar in spirit to the construction of the 
QCD moose theory in Ref.~\cite{SS}, where the holographic description
arises in the continuum limit of infinitely many hidden local symmetries
%Unlike Ref.~\cite{SS} we start directly in the continuum limit.  
(see also Ref.~\cite{Piai:2004yb}).  As in Ref.~\cite{SS}, 
vector meson dominance and QCD sum rules are natural 
consequences of our model.  Hence, the success of the model is not
coincidental, but a result of linking several proven approaches through
the AdS/CFT correspondence.
We expect the success of our model to diminish above roughly the scale given
by the mass of the lightest isospin-carrying spin-2 resonance, namely the
$a_2$ (1318 MeV \cite{PDG}).
In particular, we are completely neglecting stringy
physics which becomes important at higher energies, and we have not included 
in our description any modes with spin larger
than 1.  
At this stage, we also neglect running of the QCD coupling,  
which is likely a poor approximation for a larger range of energies.
While our model is too simple to provide a
complete dual description of QCD, its success seems to suggest that
there is a quantitatively useful reformulation of QCD as a string
theory in a higher-dimensional curved space.

\sect{Field content.}%
Table~\ref{tab:fields} illustrates the field content of our model.
The choice of the 5D fields is dictated by a principle of the 
AdS/CFT correspondence:
each operator ${\cal O}(x)$ in the 4D field
theory corresponds to a field $\phi(x,z)$ in the 5D bulk theory.
The 5D theory dual to QCD
should, therefore, contain an infinite number of fields corresponding to the
infinite number of operators in QCD.  There is, however, a small
number of operators that are important in the chiral dynamics: the
left- and right-handed currents corresponding to the SU$(N_f)_L\times$SU$(N_f)_R$
chiral flavor symmetry, and the chiral order parameter
(see Table \ref{tab:fields}). We shall include in our model only
the 5D fields which correspond to these operators and neglect all
other fields. 
\begin{table}[b]\vspace{-12pt}
\caption[]{Operators/fields of the model}\label{tab:fields}
\begin{ruledtabular}
  \begin{tabular}{p{5em}p{6em}p{3em}p{3em}p{3em}}
   4D:  ${\cal O}(x)$                  &  5D: $\phi(x,z)$  & $p$  &
$\Delta$ & $(m_5)^2$      \\
\hline
    $\bar q_L \gamma^\mu t^a q_L$ &  $A_{L\mu}^a $ & 1 & 3 & $\phantom{-}0$     \\
    $\bar q_R \gamma^\mu t^a q_R$ &  $A_{R\mu}^a$    & 1 & 3 & $\phantom{-}0$       \\
    $\overline q_R^\alpha q_L^\beta$ &  $(2/z)X^{\alpha\beta}$  & 0 &
3 & $-3$ \\
  \end{tabular}
\end{ruledtabular}
\end{table}

The 5D masses $m_5$ of the fields $A_{L\mu}^a $, $A_{R\mu}^a$, and $X$ are
determined
 via the
relation \cite{AdSCFT,AdSCFT2} $(\Delta-p)(\Delta+p-4)=m_5^2$,
where $\Delta$ is the dimension of the corresponding $p$-form 
operator---see Table~\ref{tab:fields}.  
 We assumed here that these operators keep their canonical dimensions,
 which is true only for the conserved currents. However, for the field $X$ we could easily
incorporate corrections to its classical dimension. The factor
$1/z$ in Table~\ref{tab:fields} is dictated by the dimension of the
operator $\bar q q$, while the factor of 2 is of no physical significance 
and is 
chosen for later convenience.

We shall choose the simplest possible metric for our model, 
namely, a slice of the anti-de Sitter (AdS) metric,
\begin{equation}
ds^2 =  \frac{1}{z^2}(-dz^2 + dx^\mu dx_\mu),\qquad 0<z\leq z_m.
\end{equation}
The fifth coordinate $z$ corresponds to the energy scale, as higher
energy (or momentum transfer $Q^2$) QCD physics is reflected by the behavior 
of the 
fields closer to the AdS boundary $z=0$: $Q\sim 1/z$. By virtue of the 
conformal isometry of the AdS space, in  such a model the running of
the QCD gauge coupling is neglected in a window of scales until an
infrared (IR) scale $Q_m\sim 1/z_m$.
To make the theory confining, one
introduces an IR cutoff in the metric at $z=z_m$ where spacetime
ends, in analogy with 
the case of the cascading gauge theory studied in 
Ref.~\cite{Klebanov:2000hb}.  
We shall call $z=z_m$ the ``infrared brane''  and impose
certain boundary conditions on the fields at $z=z_m$.  Certainly, this
is only a crude model of confinement.  Indeed, our model requires 
two dimensionful parameters related to chiral symmetry breaking, whereas 
in QCD there is only one. 
%(the dynamical scale $\LambdaQCD$).
%Throughout this paper, all
%integrals over $z$ are to be taken from 0 to $z_m$ unless
%otherwise specified. 
In addition, an UV cutoff can be provided by
setting the boundary to $z=\epsilon$ instead of $z=0$. Below we shall 
frequently
use such a cutoff as a mathematical tool, but we shall always imply
the limit of $\epsilon\to0$ for simplicity.
% We also frequently set the AdS curvature $R=1$.
% and measure all dimensionful parameters in terms of $R$.

\sect{5D action and chiral symmetry breaking.}The action of the
theory in the bulk is
\begin{equation}\label{5DL}
S=\int\!d^5x\,\sqrt{g}\, \tr\Bigl\{ |DX|^2  + 3 |X|^2
- \frac1{4\g^2} (F_L^2 + F_R^2)\Bigr\}
\end{equation}
where $D_\mu X=\d_\mu X - iA_{L\mu} X + iX A_{R\mu}$,
$A_{L,R}=A_{L,R}^at^a$, and %
$F_{\mu\nu} = \d_\mu A_\nu -\d_\nu A_\mu - i[A_\mu, A_\nu]$.
%$F_{L,R}=dA-i[A,A]$. 
As usual, 
%in AdS/CFT correspondence 
the gauge invariance in the 5D theory corresponds to the conservation
of the global symmetry current in the 4D theory.

At the IR brane we must impose some gauge invariant boundary conditions,
 and we make the simplest choice: $(F_L)_{z\mu}=(F_R)_{z\mu}=0$.
QCD does not {\em a priori} fix this boundary condition: for
example, 
there may be additional terms in the Lagrangian localized at $z_{m}$,
such as $|DX|^2$ and $F_{L}^2+F_R^2$.
%To some extent allowing the IR scale $z_m$ to vary in the model compensates 
%for a variation in the IR boundary condition. 
%To get a sense for the 
To estimate the
sensitivity to such terms, we checked that an 
$F^2$ boundary term with ${\cal O}(1)$ 
coefficient (keeping the $\rho$ mass fixed) yields a 10\% correction 
to the $\rho$ decay constant.  
%Hence, since we
%do not expect qualitative changes to our results from such terms, we 
%will assume for simplicity
%that such terms are absent.  
%In the following we shall be using 
We will be using
the gauge $A_z=0$ and neglecting boundary terms in the
Lagrangian. In this case our boundary conditions
are simply Neumann.

The expectation value of the field $X$ is determined by the classical
 solution satisfying the UV boundary condition $(2/\epsilon)X(\epsilon) = M$
for quark mass matrix $M$:
\begin{equation}
  X_0(z) = \frac{1}{2}M
       z + \frac{1}{2} \Sigma  z^3, \label{eq:vev}
\end{equation}
The matrix $\Sigma$ is determined by the IR boundary
condition on $X$. Instead of specifying this condition we shall choose
$\Sigma$ as an input parameter of the model. 
The meaning of $\Sigma$ in QCD can be found by
calculating the variation of the vacuum energy with respect to
$M$ \cite{Klebanov:1999tb}: 
$\Sigma^{\alpha\beta}=\<\bar q^\alpha q^\beta\>$. We shall
assume, as usual, $\Sigma=\sigma \bm1$ and take $M=m_q\bm1$.

At this stage the model has four free
parameters: $m_q$, $\sigma$, $z_m$ and $\g$.  The gauge coupling $\g$ will be
fixed by the QCD operator product expansion (OPE) for the product of currents, 
leaving three adjustable parameters.

%In this Letter 
We will focus on the $N_f=2$ lightest flavors 
%and will
%therefore restrict ourselves to SU(2)$_L \times$SU(2)$_R$ 
and neglect
effects of $O(m_q^2)$. Therefore, in Table~\ref{tab:fields},
$\alpha,\beta=1,2$; $a,b=1,2,3$ and $t^a=\sigma^a/2$, where $\sigma^a$ are
the Pauli matrices.

\sect{Matching the 5D gauge coupling.}We will use the
holographic duality to relate of the 5D coupling $\g$ in (\ref{5DL})
to the number of colors $N_c$ in QCD. The precise sense of
the holographic correspondence is the equivalence between the
generating functional of the connected correlators in the 4D theory 
$W_{4D}[\phi_0(x)]$ and the
effective action of the 5D theory $S_{5D,{\rm eff}}[\phi(x,\epsilon)]$,
with UV boundary values of the 5D bulk fields set to the value
of the sources in 4D theory: 
\begin{equation}
  W_{4D}[\phi_0(x)]=S_{5D,{\rm eff}}[\phi(x,\epsilon)]\quad\mbox{at}\quad \phi(x,\epsilon)=\phi_0(x).
\end{equation}
QCD Green's functions can therefore 
be obtained by differentiating the 5D
effective action with respect to the sources.  In the case that
stringy effects can be neglected, $S_{5D,{\rm eff}}$ is simply given by 
Eq.~(\ref{5DL}).
The action is evaluated on solutions to the 5D equations of 
motion subject to the condition that the value of each bulk field
at the boundary $z=\epsilon\to0$
%be given by a the corresponding source.
be given by the source $\phi$ of the corresponding 4D operator ${\cal
  O}$ (see Table~\ref{tab:fields}).

We may now fix the 5D gauge coupling by comparing the result for the 
vector current two-point function obtained from the above prescription
with that of QCD. Introducing the vector field as $V=(A_L+A_R)/2$, one
finds,
%to quadratic order in $V_\mu$:
%\begin{equation}
%  S= -\frac1{4\g^2}\int\!\frac{dz}z\, d^4x\, 
%F^a_V F^a_V.
%\end{equation}
in the $V_z(x,z)=0$ gauge, the equation of motion
for the transverse part of the gauge field:
\begin{equation}\label{eqVAdS}
\left[  \d_z\left(\frac1z \d_z V_\mu^a(q,z) \right) 
 + \frac{q^2}z V_\mu^a(q,z)\right]_\perp=0.
\end{equation}
Here $V_\mu^a(q,z)$ is the 4D Fourier transform of $V_\mu^a(x,z)$.  
The equations of motion are linearized, as is appropriate for determination of 
two-point functions.  
%Hadronic couplings, {\em e.g.}, $g_{\rho\pi\pi}$, depend on 
%the nonlinear dynamics.
Evaluating the action on the solution leaves only the boundary term
\begin{equation}
  S 
= -\frac 1{2\g^2}\int\!d^4x\,\left(\frac1z V_\mu^a \d_z V^{\mu a}\right)_{z=
\epsilon}.
\end{equation}
If $V_0^{\mu a}(q)$ is the Fourier transform of the source of the vector
current $J_\mu^{a}=\bar q \gamma_\mu t^a q$
at the boundary then letting $V^\mu(q,z) = V(q,z) V_0^\mu(q)$, we require that
$V(q,\epsilon)=1$.  Differentiating twice with 
respect to the source $V_0$, we arrive at the vector current 
two-point function, 
\begin{subequations}\label{Pi}
\begin{eqnarray}
  \int_x\!e^{iqx}\<J_\mu^{a}(x)J_\nu^{b}(0)\> \!&=&\! 
    \delta^{ab}(q_\mu q_\nu{-}q^2g_{\mu\nu})\Pi_{\rm V}(Q^2), \label{eq:VV}\\
  \Pi_{\rm V}(-q^2) \!&=&\! 
  \left.-\frac{1}{\g^2Q^2} \frac{\d_z V(q,z)}{z}\right|_{z=\epsilon},\label{Vqz}
\end{eqnarray}
\end{subequations} 
where $Q^2=-q^2$.
For large Euclidean $Q^2$ we only need to know $V(q,z)$ near
the boundary,
\begin{equation}\label{Bkernel}
V(Q,z)=1+\frac{Q^2z^2}{4}\ln(Q^2z^2)+ \ldots
\end{equation} 
which up to contact terms gives
\begin{equation}\label{PiVg}
  \Pi_{\rm V}(Q^2) = -\frac 1{2\g^2} \ln Q^2.
\end{equation}
On the other hand, we can compute $\Pi_V$ from QCD by evaluating
Feynman diagrams \cite{SVZ}.  The leading-order diagram is the quark bubble,
%which is dressed by gluon lines in subleading orders.  The 
%leading result is
\begin{equation}
  \Pi_{\rm V}(Q^2) = -\frac{N_c}{24\pi^2}
  %\left(1+\frac{\alpha_s}\pi\right)
     \ln Q^2.
\end{equation}
This leads to the identification
\begin{equation}
\g^2 = \frac{12 \pi^2}{N_c}\,,
\end{equation}
which completes the definition of the action (\ref{5DL}).

\sect{Hadrons.}The hadrons of QCD correspond to the
normalizable modes of the 5D fields.  These normalizable modes satisfy the
linearized equation of motion and decay sufficiently rapidly near the
boundary $z\to0$ so as to have a finite action.  
%As mentioned above, we choose Neumann boundary
%conditions at $z=z_m$, and tune $z_m$ to fit meson masses 
%(instead of
%tuning both $z_m$ and the boundary condition there).  
The IR boundary
condition gives rise to a discrete tower of normalizable modes.
The eigenvalue %of the equation of motion operator corresponding to
of a normalizable mode is the squared mass of the
corresponding meson, and the derivative of the mode near the UV boundary
yields the decay constant.

To illustrate the above, consider the tower of the $\rho$ mesons.  A
$\rho$ wavefunction, $\psi_\rho(z)$, is a solution to
Eq.~(\ref{eqVAdS}) for an arbitrary component of $V_\mu$
with $q^2=m_\rho^2$, subject to $\psi_\rho(\epsilon)=0$,
$\d_z\psi_\rho(z_m)=0$ and normalized as
$\int\!(dz/z)\, \psi_\rho(z)^2 =1$. 
Consider the Green's function corresponding to 
Eq.~(\ref{eqVAdS}) for an arbitrary component of $V^\mu$:
\begin{equation}
G(q;z,z')=\sum_\rho \frac{\psi_\rho(z)\psi_\rho(z')}{q^2-m_\rho^2+i\varepsilon}
.
\end{equation}
(The $i\varepsilon$ prescription, among other things, guarantees the positivity
of the spectral function, contrary to the claim of Ref.~\cite{Mahajan:2005rk}.)
One can show that $V(q,z')$ of Eq.~(\ref{Vqz}) is given by
$-(1/z)\d_z G(q;z,z')$ at $z = \epsilon$.  
%To improve convergence of
%the sum over the modes we add and subtract $V(0,z')$:
%$V(q,z')=-(1/z)\d_z [G(q;z,z')-G(0;z,z')]_{z=\epsilon} + V(0,z')$. 
Now from \eqref{Vqz} we find:
\begin{equation}\label{PiVrho}
\Pi_V(-q^2)= -\frac1{\g^2} \sum_\rho 
\frac{[\psi_\rho^{\ \prime}(\epsilon)/\epsilon]^2}
{(q^2-m_\rho^2+i\varepsilon)m_\rho^2}.
\end{equation} 
This allows us to extract the decay constants $F_\rho$:
\begin{equation}\label{Frho}
F_\rho^2=\frac1{\g^2}[\psi_\rho^{\ \prime}(\epsilon)/\epsilon]^2= \frac{1}{\g^2}[
\psi_\rho^{\ \prime\prime}(0)]^2,
\end{equation}
where $F_\rho$ is defined by
$\<0|J^a_\mu|\rho^b\>=F_\rho\delta^{ab}\varepsilon_\mu$ for a $\rho$ 
meson with polarization $\varepsilon_\mu$. Eqs. \eqref{PiVg} and 
\eqref{PiVrho} 
are the holographic version of the QCD sum rules.

In the axial sector ($a_1$ and $\pi$ mesons), the action to quadratic
order is
\begin{equation}
S= \int d^5x\,\left[
-\frac1{4\g^2z} F^a_A F^a_A + \frac{v(z)^2}{2z^3} (\d\pi^a-A^a)^2\right],
\end{equation}
where we have defined $v(z)=m_q z + \sigma z^3$, $A=(A_L-A_R)/2$,
and $X=X_0\exp(i2\pi^a t^a)$.  In the
$A_z=0$ gauge, the resulting equations of motion in 4D momentum space are
(%separating transverse and longitudinal modes
$A_\mu= A_{\mu\perp} + \d_\mu\varphi$)
%\newcommand{\AT}{\left(A^a_\mu\right)_\perp}
%%%
%%%  Using eqnarray here can save some space (~ 2 lines)
%%%  but makes it hard to read.
%%%
\newcommand{\AT}{A^a_\mu}
\begin{equation}
  \left[ \d_z\left(\frac1z \d_z \AT \right) + \frac{q^2}z \AT
- \frac{\g^2 v^2}{z^3} \AT\right]_\perp =0;\label{AT}
\end{equation}
\vskip -1.5em
\begin{equation}
  \d_z\left(\frac1z \d_z \varphi^a \right) 
+\frac{\g^2 v^2}{z^3} (\pi^a-\varphi^a) = 0;\label{AL}
\end{equation}
\vskip -1.5em
\begin{equation}
  -q^2\d_z\varphi^a+\frac{\g^2 v^2}{z^2} \d_z \pi^a =0.\label{Az}
\end{equation}
The $a_1$, being a spin-1 particle, is the solution to Eq.~(\ref{AT}) 
with %identical boundary conditions to the $\rho$.  
$\psi_{a_1}(0)=\d_z\psi_{a_1}(z_m)=0$.
The $a_1$ decay constant, $F_{a_1}$, is given by an expression similar to 
Eq.~(\ref{Frho}), but with $\rho$ replaced by ${a_1}$.

Our theory has all the consequences of chiral symmetry built in.
%For instance, 
%the apparent invariance of the action, and
%consequently, of the equations w.r.t. a $z$-independent gauge
%transformation $\pi\to\pi+\alpha(x)$ and $\varphi\to\varphi+\alpha(x)$,
%corresponds to the operator  identity 
%$\d_\mu J_5^{\mu a} = 2im_q \bar q \gamma_5 t^a q$.
%More importantly for our purposes,
Let us derive the Gell-Mann--Oakes--Renner (GOR) relation, 
\begin{equation}
  m_\pi^2f_\pi^2=(m_u+m_d)\<\bar{q}q\>=2m_q\sigma.
\end{equation}  
Since
$\<0|A_\mu|\pi\>=i f_\pi q_\mu$, the 
axial current correlator in the $m_\pi=0$ limit has 
a singularity at $q^2=0$:
$\Pi_{\rm A}(-q^2)\to-f_\pi^2 /q^2$.
Using the holographic recipe [cf. Eq.~(\ref{Pi})],
\begin{equation}\label{fpi}
f_\pi^2 = -\frac1{\g^2}\left.\frac{\d_z A(0,z)}{z}\right|_{z=\epsilon},
\end{equation}
where $A(0,z)$ is the solution to Eq.~(\ref{AT}) with $q^2=0$,
satisfying $A'(0,z_m)=0$, $A(0,\epsilon)=1$.
The pion is the solution to Eqs. 
\eqref{AL} and \eqref{Az}, 
subject to $\varphi'(z_m)=\varphi(\epsilon)=\pi(\epsilon)=0$.
We may construct such a solution perturbatively in $m_\pi$ by letting
$\varphi(z)=A(0,z)-1$. Then, from Eq.~(\ref{Az}), to leading order in 
$m_\pi^2$,
\begin{equation}
\pi(z) = {m_\pi^2}\int^z_0\! du 
\frac{u^3}{v(u)^2}\frac1{\g^2u}\d_u A(0,u)\,.
\end{equation}
The function $u^3/v(u)^2$ has a significant support only for
$u\sim z_c\equiv\sqrt{m_q/\sigma}$. The function  $\d_u A(0,u)/(\g^2u)$
for such small values of $u$ can be replaced by its value at 
$u=\epsilon$, which is related to $f_\pi$ via
(\ref{fpi}). Performing the integral
one finds that $\pi=-m_\pi^2 f_\pi^2/(2m_q\sigma)$ for $z\gg z_c$.
%for all $z$ except a small region $0<z\lesssim z_c$ over 
%which it quickly rises
%from $\pi(0)=0$. 
Equations (\ref{AT}) and (\ref{AL}) are solved by
$\varphi=A(0,z)-1$ 
and $\pi={\rm const}$ 
for $z\gg z_c$ only if $\pi=-1$, hence 
$m_\pi^2 f_\pi^2=2m_q\sigma+{\cal O}(m_q^2)$.

\sect{Meson interactions and $g_{\rho\pi\pi}$.}%
The meson interactions
can be read from the nonbilinear terms in the 5D
action.
%\begin{equation}
%S_{\pi\rho}= 
%\int d^5x\,\frac{v(z)^2}{2z^3} (\d\pi^a-A^a+\epsilon^{abc}V_b\pi_c)^2.
%\end{equation}
For example, we find that the $\pi$-$\rho$ coupling is given by
%\begin{equation}\label{grpp}
%g_{\rho\pi\pi} = \g \int\! dz\, \frac{v^2(z)}{z^3}\, 
%(\pi-\varphi)(z)\, \pi(z)\, \psi_\rho(z),
%\end{equation}
\begin{equation}\label{grpp}
g_{\rho\pi\pi} = \g \int\! dz\, \psi_\rho(z)
\left(\frac{\phi'(z)^2}{\g^2\, z} + \frac{v(z)^2 (\pi-\phi)^2}{z^3} \right).
\end{equation}
The normalization of $\pi$ is fixed by the pion kinetic term:
integrating the function in parentheses in Eq.~\eqref{grpp} gives~1.
%The canonical normalization of the pion kinetic term translates into
%integral of the function in parenthesis is equal to one
%where the pion wavefunction $\pi$ is normalized such that
%the pion kinetic term is canonically normalized. We
%have not included operators like $F_{L,R}^3$ in the action, although they may 
%not be negligible on the rho mass shell.
One must be aware that this 3-meson amplitude could be
sensitive to the $F^3$ terms not yet included in our model.

\sect{Predictions.}%
%The model described above in some ways mimics the 
%approach of QCD sum rules, and we naturally expect our model to work about as
%well, at roughly the 20\% level. 
%\vspace{-6pt}
%In our numerics we set the UV cutoff $\epsilon=10^{-10}$ MeV$^{-1}$ and 
%checked that this value gives numerical results well within 0.1\% 
%of the $\epsilon\rightarrow0$ limit. 
From Eq.~(\ref{eqVAdS}) and the Dirichlet boundary conditions, 
the $\rho$ wavefunctions are Bessel functions
with masses determined by zeroes of $J_0(q z_m)$.  Hence,
%The $\rho$ wavefunction is the lowest resonance satisfying (\ref{eqVAdS}).
%The solutions to (\ref{eqVAdS}) are Bessel functions and 
%as the UV cutoff $\epsilon$ is taken to zero, 
%the eigenstates
%have masses determined by zeroes of $J_0(q z_m)$.  Hence, 
$m_\rho=2.405/z_m=776$ MeV fixes
$z_m=1/(323\ {\rm MeV})$.   
$m_q$ and $\sigma$ can then be fit to the experimental values of $m_\pi$ and $f_\pi$, yielding
$m_q=2.29$ MeV and $\sigma=(327\ {\rm MeV})^3$. 
These parameters correspond to Model A in Table~\ref{Table}.

% As a measure of the success of our
%model we calculate the root-mean-square (rms) 
%fractional error in the observables $O_i$, defined by, 
%\begin{equation} 
%E_n\equiv\sqrt{\frac{1}{n}\,\sum_i \frac{(O_i(measured)-O_i(model))^2}{O_i(measured)^2}},
%\end{equation}
%where the number of degrees of freedom, $n$, 
%is the difference between the number of observables and number of
%model parameters.

%The experimental value of the rho mass, $m_\rho=775.8\pm0.5$ 
%MeV \cite{PDG}, then fixes $z_m=1/(323\ {\rm MeV})$.
%The remaining parameters, $m_q$ and $\sigma$, are fit to the experimental values
%$m_\pi=139.6\pm0.0004$ MeV  \cite{PDG} and $f_{\pi}=92.4\pm0.35$ MeV \cite{PDG}, 
%yielding
%$m_q=2.29$ MeV and $\sigma=(327\ {\rm MeV})^3$.  
%(We quote data for the charged mesons and ignore
%electromagnetic corrections, which are typically of order a few percent.)
%With these values of the  parameters (Model A), the third
%column of Table~\ref{Table} lists our calculated observables.

%the remaining four observables which we have
%calculated in this model are
%given in the third column of Table~\ref{Table}. %%  For the 
%% measured value of $f_{a_1}$ quoted in the second column, 
%% which has relatively
%% large error bars, we took the average value obtained from lattice data and tau 
%% observables,
%% and we added the corresponding experimental
%% errors by quadratures \cite{SS}. 
%As a measure of the success of
%our model we used the rms error, 
The rms error, 
$\varepsilon_{\rm rms}=\left(\sum_O (\delta O/O)^2/n\right)^{1/2}$ 
(where $\delta O/O$ is the fractional error of an observable $O$ and
$n=4$ equals the number of observables minus the number of parameters)
%The rms error 
for Model A is 15\%.  

%We also performed a global fit of the model to the central values of all seven
%observables simultaneously by minimizing the rms error.
%The best fit is given by
A global fit to all seven observables (Model B) yields the parameters,
$z_m=1/(346\ {\rm MeV})$, $m_q=2.3$ MeV and $\sigma=(308\ {\rm MeV})^3$.  
The last column of Table~\ref{Table} lists the calculated observables in
this model.
%The
%values of the observables with this fit (Model B) are shown in the
%last column of Table~\ref{Table}.  
The rms error of Model B
is a remarkably small~9\%.

\sect{Discussion and outlook.}The holographic model of QCD studied here
is quite crude and 
depends on only three free parameters,
but it agrees surprisingly well with the seven experimentally
measured observables which we have studied.
There are several ways in which we may attempt to extend and improve
the model.  
(i)
The glueball spectrum
can be calculated from the gravitational and dilaton
modes in the theory, which were not included in this study.
(ii)
It is straightforward to describe power corrections in the current
correlators \cite{ADSOPE}.  
Here we matched the gauge coupling $\g$
in our model to the leading term---the unit operator---in the OPE 
of the product of currents.
Higher dimension operators also appear in the 
OPE, suppressed by powers of the Euclidean momentum $Q$.
These corrections can be calculated in QCD \cite{SVZ}.
In the holographic model, these corrections arise from trilinear
and higher terms in the 5D action, such as $\int d^5x\,\sqrt{g} X^2 F^2$.
Matching the QCD OPE coefficients to the coefficients of the 5D action
provides a method of building and constraining the effective 5D action. 
(iii)
Including the strange quark
into the model with an approximate SU(3)$\times$SU(3) chiral symmetry
is a natural extension of the model.
(iv)
The chiral anomaly can be incorporated via a 5D Chern-Simons term.
(v) We can include corrections to the
dimensions of the chiral order parameters by varying the mass of 
the corresponding
fields $X$ in the 5D theory, and we can include running of the gauge coupling
via logarithmic corrections to the
AdS geometry.
It is interesting to note in this context, that those results 
which follow from partial
conservation of the axial current,  e.g., the GOR relation, 
continue to hold as we vary the 5D mass of $X$ in the model~\cite{ADSOPE}.

\begin{table}[t]\vspace{-6pt}
\caption[]{Results of the model for QCD observables.  Model A is a fit of the 
three
model parameters to $m_\pi$, $f_\pi$ and $m_\rho$ (see asterisks).  
Model B is a fit to all
seven observables.}
%\smallskip
\begin{ruledtabular}
\begin{tabular}{cccc}  
                   & Measured                             & Model A        &Model B\\
Observable         & (MeV)                                & (MeV)          & (MeV) \\ \hline
$m_\pi$            & 139.6$\pm0.0004$ \cite{PDG}          & 139.6$^*$          & 141 \\
$m_\rho$           & 775.8$\pm 0.5$   \cite{PDG}          & 775.8$^*$          & 832   \\ 
$m_{a_1}$          & 1230$\pm40$      \cite{PDG}          & 1363           & 1220  \\
$f_\pi$            & 92.4$\pm0.35$    \cite{PDG}          & 92.4$^*$        & 84.0  \\
$F_\rho^{\,1/2}$   & 345$\pm8$        \cite{Donoghue}     & 329            & 353   \\
$F_{a_1}^{\,1/2}$  & 433$\pm13$     \cite{SS,Isgur:1988vm}& 486            & 440   \\
$g_{\rho\pi\pi}$   & 6.03$\pm0.07$    \cite{PDG}          & 4.48           & 5.29  \\ 
\end{tabular}\vspace{-12pt}
\end{ruledtabular}
\label{Table}
\end{table}

%\sect{Acknowledgments}
This work
is supported in part by DOE Grants DE-FG0201-ER41195 and DE-AC02-76SF00515. 
  J.E. thanks the College of William and Mary for support.
D.T.S. and M.A.S. thank A.P.~Sloan Foundation for support.

\end{document}